\documentclass[letterpaper,compsoc,twoside]{IEEEtran}
\usepackage{fixltx2e} 
\usepackage{cmap} 
\usepackage{ifthen}
\usepackage[T1]{fontenc}
\usepackage[utf8]{inputenc}
\usepackage{amsmath}

\usepackage[font={small,it},labelfont=bf]{caption}
\usepackage{float}

\usepackage{textcomp} 

\pdfoutput=1
\usepackage{scipy}
\makeatletter
\def\PY@reset{\let\PY@it=\relax \let\PY@bf=\relax%
    \let\PY@ul=\relax \let\PY@tc=\relax%
    \let\PY@bc=\relax \let\PY@ff=\relax}
\def\PY@tok#1{\csname PY@tok@#1\endcsname}
\def\PY@toks#1+{\ifx\relax#1\empty\else%
    \PY@tok{#1}\expandafter\PY@toks\fi}
\def\PY@do#1{\PY@bc{\PY@tc{\PY@ul{%
    \PY@it{\PY@bf{\PY@ff{#1}}}}}}}
\def\PY#1#2{\PY@reset\PY@toks#1+\relax+\PY@do{#2}}

\expandafter\def\csname PY@tok@gd\endcsname{\def\PY@tc##1{\textcolor[rgb]{0.63,0.00,0.00}{##1}}}
\expandafter\def\csname PY@tok@gu\endcsname{\let\PY@bf=\textbf\def\PY@tc##1{\textcolor[rgb]{0.50,0.00,0.50}{##1}}}
\expandafter\def\csname PY@tok@gt\endcsname{\def\PY@tc##1{\textcolor[rgb]{0.00,0.27,0.87}{##1}}}
\expandafter\def\csname PY@tok@gs\endcsname{\let\PY@bf=\textbf}
\expandafter\def\csname PY@tok@gr\endcsname{\def\PY@tc##1{\textcolor[rgb]{1.00,0.00,0.00}{##1}}}
\expandafter\def\csname PY@tok@cm\endcsname{\let\PY@it=\textit\def\PY@tc##1{\textcolor[rgb]{0.25,0.50,0.56}{##1}}}
\expandafter\def\csname PY@tok@vg\endcsname{\def\PY@tc##1{\textcolor[rgb]{0.73,0.38,0.84}{##1}}}
\expandafter\def\csname PY@tok@vi\endcsname{\def\PY@tc##1{\textcolor[rgb]{0.73,0.38,0.84}{##1}}}
\expandafter\def\csname PY@tok@mh\endcsname{\def\PY@tc##1{\textcolor[rgb]{0.13,0.50,0.31}{##1}}}
\expandafter\def\csname PY@tok@cs\endcsname{\def\PY@tc##1{\textcolor[rgb]{0.25,0.50,0.56}{##1}}\def\PY@bc##1{\setlength{\fboxsep}{0pt}\colorbox[rgb]{1.00,0.94,0.94}{\strut ##1}}}
\expandafter\def\csname PY@tok@ge\endcsname{\let\PY@it=\textit}
\expandafter\def\csname PY@tok@vc\endcsname{\def\PY@tc##1{\textcolor[rgb]{0.73,0.38,0.84}{##1}}}
\expandafter\def\csname PY@tok@il\endcsname{\def\PY@tc##1{\textcolor[rgb]{0.13,0.50,0.31}{##1}}}
\expandafter\def\csname PY@tok@go\endcsname{\def\PY@tc##1{\textcolor[rgb]{0.20,0.20,0.20}{##1}}}
\expandafter\def\csname PY@tok@cp\endcsname{\def\PY@tc##1{\textcolor[rgb]{0.00,0.44,0.13}{##1}}}
\expandafter\def\csname PY@tok@gi\endcsname{\def\PY@tc##1{\textcolor[rgb]{0.00,0.63,0.00}{##1}}}
\expandafter\def\csname PY@tok@gh\endcsname{\let\PY@bf=\textbf\def\PY@tc##1{\textcolor[rgb]{0.00,0.00,0.50}{##1}}}
\expandafter\def\csname PY@tok@ni\endcsname{\let\PY@bf=\textbf\def\PY@tc##1{\textcolor[rgb]{0.84,0.33,0.22}{##1}}}
\expandafter\def\csname PY@tok@nl\endcsname{\let\PY@bf=\textbf\def\PY@tc##1{\textcolor[rgb]{0.00,0.13,0.44}{##1}}}
\expandafter\def\csname PY@tok@nn\endcsname{\let\PY@bf=\textbf\def\PY@tc##1{\textcolor[rgb]{0.05,0.52,0.71}{##1}}}
\expandafter\def\csname PY@tok@no\endcsname{\def\PY@tc##1{\textcolor[rgb]{0.38,0.68,0.84}{##1}}}
\expandafter\def\csname PY@tok@na\endcsname{\def\PY@tc##1{\textcolor[rgb]{0.25,0.44,0.63}{##1}}}
\expandafter\def\csname PY@tok@nb\endcsname{\def\PY@tc##1{\textcolor[rgb]{0.00,0.44,0.13}{##1}}}
\expandafter\def\csname PY@tok@nc\endcsname{\let\PY@bf=\textbf\def\PY@tc##1{\textcolor[rgb]{0.05,0.52,0.71}{##1}}}
\expandafter\def\csname PY@tok@nd\endcsname{\let\PY@bf=\textbf\def\PY@tc##1{\textcolor[rgb]{0.33,0.33,0.33}{##1}}}
\expandafter\def\csname PY@tok@ne\endcsname{\def\PY@tc##1{\textcolor[rgb]{0.00,0.44,0.13}{##1}}}
\expandafter\def\csname PY@tok@nf\endcsname{\def\PY@tc##1{\textcolor[rgb]{0.02,0.16,0.49}{##1}}}
\expandafter\def\csname PY@tok@si\endcsname{\let\PY@it=\textit\def\PY@tc##1{\textcolor[rgb]{0.44,0.63,0.82}{##1}}}
\expandafter\def\csname PY@tok@s2\endcsname{\def\PY@tc##1{\textcolor[rgb]{0.25,0.44,0.63}{##1}}}
\expandafter\def\csname PY@tok@nt\endcsname{\let\PY@bf=\textbf\def\PY@tc##1{\textcolor[rgb]{0.02,0.16,0.45}{##1}}}
\expandafter\def\csname PY@tok@nv\endcsname{\def\PY@tc##1{\textcolor[rgb]{0.73,0.38,0.84}{##1}}}
\expandafter\def\csname PY@tok@s1\endcsname{\def\PY@tc##1{\textcolor[rgb]{0.25,0.44,0.63}{##1}}}
\expandafter\def\csname PY@tok@ch\endcsname{\let\PY@it=\textit\def\PY@tc##1{\textcolor[rgb]{0.25,0.50,0.56}{##1}}}
\expandafter\def\csname PY@tok@m\endcsname{\def\PY@tc##1{\textcolor[rgb]{0.13,0.50,0.31}{##1}}}
\expandafter\def\csname PY@tok@gp\endcsname{\let\PY@bf=\textbf\def\PY@tc##1{\textcolor[rgb]{0.78,0.36,0.04}{##1}}}
\expandafter\def\csname PY@tok@sh\endcsname{\def\PY@tc##1{\textcolor[rgb]{0.25,0.44,0.63}{##1}}}
\expandafter\def\csname PY@tok@ow\endcsname{\let\PY@bf=\textbf\def\PY@tc##1{\textcolor[rgb]{0.00,0.44,0.13}{##1}}}
\expandafter\def\csname PY@tok@sx\endcsname{\def\PY@tc##1{\textcolor[rgb]{0.78,0.36,0.04}{##1}}}
\expandafter\def\csname PY@tok@bp\endcsname{\def\PY@tc##1{\textcolor[rgb]{0.00,0.44,0.13}{##1}}}
\expandafter\def\csname PY@tok@c1\endcsname{\let\PY@it=\textit\def\PY@tc##1{\textcolor[rgb]{0.25,0.50,0.56}{##1}}}
\expandafter\def\csname PY@tok@o\endcsname{\def\PY@tc##1{\textcolor[rgb]{0.40,0.40,0.40}{##1}}}
\expandafter\def\csname PY@tok@kc\endcsname{\let\PY@bf=\textbf\def\PY@tc##1{\textcolor[rgb]{0.00,0.44,0.13}{##1}}}
\expandafter\def\csname PY@tok@c\endcsname{\let\PY@it=\textit\def\PY@tc##1{\textcolor[rgb]{0.25,0.50,0.56}{##1}}}
\expandafter\def\csname PY@tok@mf\endcsname{\def\PY@tc##1{\textcolor[rgb]{0.13,0.50,0.31}{##1}}}
\expandafter\def\csname PY@tok@err\endcsname{\def\PY@bc##1{\setlength{\fboxsep}{0pt}\fcolorbox[rgb]{1.00,0.00,0.00}{1,1,1}{\strut ##1}}}
\expandafter\def\csname PY@tok@mb\endcsname{\def\PY@tc##1{\textcolor[rgb]{0.13,0.50,0.31}{##1}}}
\expandafter\def\csname PY@tok@ss\endcsname{\def\PY@tc##1{\textcolor[rgb]{0.32,0.47,0.09}{##1}}}
\expandafter\def\csname PY@tok@sr\endcsname{\def\PY@tc##1{\textcolor[rgb]{0.14,0.33,0.53}{##1}}}
\expandafter\def\csname PY@tok@mo\endcsname{\def\PY@tc##1{\textcolor[rgb]{0.13,0.50,0.31}{##1}}}
\expandafter\def\csname PY@tok@kd\endcsname{\let\PY@bf=\textbf\def\PY@tc##1{\textcolor[rgb]{0.00,0.44,0.13}{##1}}}
\expandafter\def\csname PY@tok@mi\endcsname{\def\PY@tc##1{\textcolor[rgb]{0.13,0.50,0.31}{##1}}}
\expandafter\def\csname PY@tok@kn\endcsname{\let\PY@bf=\textbf\def\PY@tc##1{\textcolor[rgb]{0.00,0.44,0.13}{##1}}}
\expandafter\def\csname PY@tok@cpf\endcsname{\let\PY@it=\textit\def\PY@tc##1{\textcolor[rgb]{0.25,0.50,0.56}{##1}}}
\expandafter\def\csname PY@tok@kr\endcsname{\let\PY@bf=\textbf\def\PY@tc##1{\textcolor[rgb]{0.00,0.44,0.13}{##1}}}
\expandafter\def\csname PY@tok@s\endcsname{\def\PY@tc##1{\textcolor[rgb]{0.25,0.44,0.63}{##1}}}
\expandafter\def\csname PY@tok@kp\endcsname{\def\PY@tc##1{\textcolor[rgb]{0.00,0.44,0.13}{##1}}}
\expandafter\def\csname PY@tok@w\endcsname{\def\PY@tc##1{\textcolor[rgb]{0.73,0.73,0.73}{##1}}}
\expandafter\def\csname PY@tok@kt\endcsname{\def\PY@tc##1{\textcolor[rgb]{0.56,0.13,0.00}{##1}}}
\expandafter\def\csname PY@tok@sc\endcsname{\def\PY@tc##1{\textcolor[rgb]{0.25,0.44,0.63}{##1}}}
\expandafter\def\csname PY@tok@sb\endcsname{\def\PY@tc##1{\textcolor[rgb]{0.25,0.44,0.63}{##1}}}
\expandafter\def\csname PY@tok@k\endcsname{\let\PY@bf=\textbf\def\PY@tc##1{\textcolor[rgb]{0.00,0.44,0.13}{##1}}}
\expandafter\def\csname PY@tok@se\endcsname{\let\PY@bf=\textbf\def\PY@tc##1{\textcolor[rgb]{0.25,0.44,0.63}{##1}}}
\expandafter\def\csname PY@tok@sd\endcsname{\let\PY@it=\textit\def\PY@tc##1{\textcolor[rgb]{0.25,0.44,0.63}{##1}}}

\def\PYZus{\char`\_}

\def\PYZhy{\char`\-}
\def\PYZsq{\char`\'}

\makeatother

\providecommand*{\DUfootnotemark}[3]{%
  \raisebox{1em}{\hypertarget{#1}{}}%
  \hyperlink{#2}{\textsuperscript{#3}}%
}
\providecommand{\DUfootnotetext}[4]{%
  \begingroup%
  \renewcommand{\thefootnote}{%
    \protect\raisebox{1em}{\protect\hypertarget{#1}{}}%
    \protect\hyperlink{#2}{#3}}%
  \footnotetext{#4}%
  \endgroup%
}

\providecommand*{\DUrole}[2]{%
  \ifcsname DUrole#1\endcsname%
    \csname DUrole#1\endcsname{#2}%
  \else
    \ifcsname docutilsrole#1\endcsname%
      \csname docutilsrole#1\endcsname{#2}%
    \else%
      #2%
    \fi%
  \fi%
}

\ifthenelse{\isundefined{\hypersetup}}{
  \usepackage[colorlinks=true,linkcolor=blue,urlcolor=blue]{hyperref}
  \urlstyle{same} 
}{}

\begin{document}
\newcounter{footnotecounter}\title{Plyades: A Python Library for Space Mission Design}\author{Helge Eichhorn$^{\setcounter{footnotecounter}{1}\fnsymbol{footnotecounter}\setcounter{footnotecounter}{2}\fnsymbol{footnotecounter}}$%
          \setcounter{footnotecounter}{1}\thanks{\fnsymbol{footnotecounter} %
          Corresponding author: \protect\href{mailto:eichhorn@dik.tu-darmstadt.de}{eichhorn@dik.tu-darmstadt.de}}\setcounter{footnotecounter}{2}\thanks{\fnsymbol{footnotecounter} Technische Universität Darmstadt, Department of Computer Integrated Design}, Reiner Anderl$^{\setcounter{footnotecounter}{2}\fnsymbol{footnotecounter}}$\thanks{%

          \noindent%
          Copyright\,\copyright\,2015 Helge Eichhorn et al. This is an open-access article distributed under the terms of the Creative Commons Attribution License, which permits unrestricted use, distribution, and reproduction in any medium, provided the original author and source are credited. http://creativecommons.org/licenses/by/3.0/%
        }}\maketitle
          \renewcommand{\leftmark}{PROC. OF THE 8th EUR. CONF. ON PYTHON IN SCIENCE (EUROSCIPY 2015)}
          \renewcommand{\rightmark}{PLYADES: A PYTHON LIBRARY FOR SPACE MISSION DESIGN}

\setcounter{page}{9}
\newcommand*{\docutilsroleref}{\ref}
\newcommand*{\docutilsrolelabel}{\label}
\AtEndDocument{\cleardoublepage}
\begin{abstract}Designing a space mission is a computation-heavy task.
Software tools that conduct the necessary numerical simulations and optimizations are therefore indispensable.
The usability of existing software, written in Fortran and MATLAB, suffers because of high complexity, low levels of abstraction and out-dated programming practices.
We propose Python as a viable alternative for astrodynamics tools and demonstrate the proof-of-concept library Plyades which combines powerful features with Pythonic ease of use.\end{abstract}\begin{IEEEkeywords}data modeling, object-oriented programming, orbital mechanics, astrodynamics\end{IEEEkeywords}

\section{Introduction%
  \label{introduction}%
}

Designing a space mission trajectory is a computation-heavy task.
Software tools that conduct the necessary numerical simulations and optimizations are therefore indispensable and high numerical performance is required.
No science mission or spacecraft are exactly the same and during the early mission design phases the technical capabilities and constraints change frequently.
Therefore high development speed and programmer productivity are required as well.
Due to its supreme numerical performance Fortran has been the top programming language in many organizations within the astrodynamics community.

At the European Space Operations Center (ESOC) of the European Space Agency (ESA) a large body of sometimes decades old Fortran77 code remains in use for mission analysis tasks.
While this legacy code mostly fulfills the performance requirements\DUfootnotemark{id1}{id3}{1} usability and programmer productivity suffer.
One reason for this is that Fortran is a compiled language which hinders exploratory analyses and rapid prototyping.
The low level of abstraction supported by Fortran77 and programming conventions, like the 6-character variable name limit and fixed-form source code, that are in conflict with today's best practices are a more serious problem, though.
The more recent Fortran standards remedy a lot of these shortcomings, e.g. free-form source in Fortran90 or object-oriented programming features in Fortran2003, but also introduce new complexity, e.g. requiring sophisticated build systems for dependency resolution.
Compiler vendors have also been very slow to implement new standards.
For example this year the Intel Fortran compiler achieved full support of the Fortran2003 standard, which was released in 2005 \cite{IFC15}.%
\DUfootnotetext{id3}{id1}{1}{
Many routines were not written with thread-safety and re-entrancy in mind and can therefore not be used in parallel codes.}

Due to these reasons Fortran-based tools and libraries have been generally used together with programming environments with better usability such as MATLAB.
A common approach for developing mission design software at ESOC is prototyping and implementing downstream processes such as visualization in MATLAB and then later porting performance-intensive parts or the whole system to Fortran77.
The results are added complexity through the use of the MEX-interface for integrating Fortran and MATLAB, duplicated effort for porting, and still a low-level of abstraction because the system design is constrained by Fortran's limitations.

Because of the aforementioned problems some organizations explore possibilities to replace Fortran for future development.
The French space agency CNES (Centre National D'Études Spatiales) for instance uses the Java-based Orekit library \cite{Ore15} for its flight dynamics systems.

In this paper we show why Python and the scientific Python ecosystem are a viable choice for the next generation of space mission design software and present the Plyades library.
Plyades is a proof-of-concept implementation of an object-oriented astrodynamics library in pure Python.
It makes use of many prominent scientific Python libraries such as Numpy, Scipy, Matplotlib, Bokeh, and Astropy.
In the following we discuss the design of the Plyades data model and conclude the paper with an exemplary analysis.

\section{Why Python?%
  \label{why-python}%
}

Perez, Granger, and Hunter \cite{PGH11} show that the scientific Python ecosystem has reached a high level of maturity and conclude that \textquotedbl{}Python has now entered a phase where it's clearly a valid choice for high-level scientific code development, and its use is rapidly growing\textquotedbl{}.
This assessment also holds true for astrodynamics work as most of the required low-level mathematical and infrastructural building blocks are already available, as shown below:%
\begin{itemize}

\item 

Vector algebra: NumPy \cite{WCV11}
\item 

Visualization: Matplotlib \cite{JDH07}, Bokeh \cite{BDT15}
\item 

Numerical integration and optimization: SciPy \cite{JOP01}
\item 

High performance numerics: Cython \cite{BBC11}, Numba \cite{NDT15}
\item 

Planetary ephemerides: jplephem \cite{BRh15}
\end{itemize}

Another advantage is Python's friendliness to beginners.
In the United States Python was the most popular language for teaching introductory computer science (CS) courses at top-ranked CS-departments in 2014 \cite{PGu14}.
Astrodynamicists are rarely computer scientists but mostly aerospace engineers, physicists and mathematicians.
Most graduates of these disciplines have only little programming experience.
Moving to Python could therefore lower the barrier of entry significantly.

It is also beneficial that the scientific Python ecosystem is open-source compared to the MATLAB environment and commercial Fortran compilers which require expensive licenses.

\section{Design of the Plyades Object Model%
  \label{design-of-the-plyades-object-model}%
}

The general idea behind the design of the Plyades library is the introduction of proper abstraction.
We developed a domain model based on the following entities which are part of every analysis in mission design:%
\begin{itemize}

\item 

orbits (or trajectories),
\item 

spacecraft state vectors,
\item 

and celestial bodies.
\end{itemize}

\subsection{The Body Class%
  \label{the-body-class}%
}

The Body class is a simple helper class that holds physical constants and other properties of celestial bodies such as planets and moons.
These include%
\begin{itemize}

\item 

the name of the body,
\item 

the gravitational parameter $\mu$,
\item 

the mean radius $r_m$,
\item 

the equatorial radius $r_e$,
\item 

the polar radius $r_p$,
\item 

the $J_2$ coefficient of the body's gravity potential.
\end{itemize}


\subsection{The State Class%
  \label{the-state-class}%
}

To define the state of a spacecraft in space-time we need the following information%
\begin{itemize}

\item 

the position vector ($\vec{r} \in \mathbf{r^3}$),
\item 

the velocity vector ($\vec{v} \in \mathbf{r^3}$),
\item 

the corresponding moment in time, the so-called epoch,
\item 

the reference frame in which the vectors are defined,
\item 

the central body which also defines the origin of the coordinate system,
\item 

and additional spacecraft status parameters such as mass.
\end{itemize}

While the information could certainly be stored in a single Numpy-array an object-oriented programming (OOP) approach offers advantages.
Since all necessary data can be encapsulated in the object most orbital characteristics can be calculated by calling niladic or monadic instance methods.
Keeping the number of parameters within the application programming interface (API) very small, as recommended by Robert C. Martin \cite{RCM08}, improves usability, e.g. the user is not required to know the order of the function parameters.
OOP also offers the opportunity to integrate the \texttt{State} class with the Python object model and the Jupyter notebook to provide rich human-friendly representations.

State vectors also provide methods for backwards and forwards propagation.
Through propagation trajectories are generated, which are instances of the \texttt{Orbit} class.

\subsection{The Orbit Class%
  \label{the-orbit-class}%
}

In contrast to the \texttt{State} class which represents a single state in space-time the \texttt{Orbit} class spans a time interval and contains several spacecraft states.
It provides all necessary tools to analyze the evolution of the trajectory over time including%
\begin{itemize}

\item 

quick visualizations in three-dimensional space and two-dimensional projections,
\item 

evolution of orbital characteristics,
\item 

and determination of intermediate state vectors.
\end{itemize}

\section{Exemplary Usage%
  \label{exemplary-usage}%
}

In this example we use the Plyades library to conduct an analysis of the orbit of the International Space Station (ISS)\DUfootnotemark{id15}{id17}{2}.
We obtain the inital state data on August 28, 2015, 12:00h from NASA realtime trajectory data \cite{NAS15} and  use it to instantiate a Plyades \texttt{State} object as shown below.%
\DUfootnotetext{id17}{id15}{2}{
A Jupyter notebook with this analysis can be obtained from \href{https://github.com/helgee/euroscipy-2015}{Github}.}
\begin{Verbatim}[commandchars=\\\{\},fontsize=\footnotesize]
\PY{n}{iss\PYZus{}r} \PY{o}{=} \PY{n}{numpy}\PY{o}{.}\PY{n}{array}\PY{p}{(}\PY{p}{[}
    \PY{o}{\PYZhy{}}\PY{l+m+mf}{2775.03475}\PY{p}{,}
    \PY{l+m+mf}{4524.24941}\PY{p}{,}
    \PY{l+m+mf}{4207.43331}\PY{p}{,}
    \PY{p}{]}\PY{p}{)} \PY{o}{*} \PY{n}{astropy}\PY{o}{.}\PY{n}{units}\PY{o}{.}\PY{n}{km}
\PY{n}{iss\PYZus{}v} \PY{o}{=} \PY{n}{numpy}\PY{o}{.}\PY{n}{array}\PY{p}{(}\PY{p}{[}
    \PY{o}{\PYZhy{}}\PY{l+m+mf}{3.641793088}\PY{p}{,}
    \PY{o}{\PYZhy{}}\PY{l+m+mf}{5.665088604}\PY{p}{,}
    \PY{l+m+mf}{3.679500667}\PY{p}{,}
    \PY{p}{]}\PY{p}{)} \PY{o}{*} \PY{n}{astropy}\PY{o}{.}\PY{n}{units}\PY{o}{.}\PY{n}{km}\PY{o}{/}\PY{n}{astropy}\PY{o}{.}\PY{n}{units}\PY{o}{.}\PY{n}{s}
\PY{n}{iss\PYZus{}t} \PY{o}{=} \PY{n}{astropy}\PY{o}{.}\PY{n}{time}\PY{o}{.}\PY{n}{Time}\PY{p}{(}\PY{l+s+s1}{\PYZsq{}}\PY{l+s+s1}{2015\PYZhy{}08\PYZhy{}28T12:00:00.000}\PY{l+s+s1}{\PYZsq{}}\PY{p}{)}
\PY{n}{frame} \PY{o}{=} \PY{l+s+s1}{\PYZsq{}}\PY{l+s+s1}{ECI}\PY{l+s+s1}{\PYZsq{}}
\PY{n}{body} \PY{o}{=} \PY{n}{plyades}\PY{o}{.}\PY{n}{bodies}\PY{o}{.}\PY{n}{EARTH}

\PY{n}{iss} \PY{o}{=} \PY{n}{plyades}\PY{o}{.}\PY{n}{State}\PY{p}{(}\PY{n}{iss\PYZus{}r}\PY{p}{,} \PY{n}{iss\PYZus{}v}\PY{p}{,} \PY{n}{iss\PYZus{}t}\PY{p}{,} \PY{n}{frame}\PY{p}{,} \PY{n}{body}\PY{p}{)}
\end{Verbatim}
The position (\texttt{iss\_r}) and velocity (\texttt{iss\_v}) vectors use the units functionality from the Astropy package \cite{ASP13} while the timestamp (\texttt{iss\_t}) is an Astropy \texttt{Time} object.
The constant \texttt{EARTH} from the \texttt{plyades.bodies} module is a \texttt{Body} object and provides Earth's planetary constants.

The resulting \texttt{State} object contains all data necessary to describe the current orbit of the spacecraft.
Calculations of orbital characteristics are therefore implemented with the \texttt{@property} decorator, like shown below, and are instantly available.\begin{Verbatim}[commandchars=\\\{\},fontsize=\footnotesize]
\PY{n+nd}{@property}
\PY{k}{def} \PY{n+nf}{elements}\PY{p}{(}\PY{n+nb+bp}{self}\PY{p}{)}\PY{p}{:}
    \PY{k}{return} \PY{n}{kepler}\PY{o}{.}\PY{n}{elements}\PY{p}{(}\PY{n+nb+bp}{self}\PY{o}{.}\PY{n}{body}\PY{o}{.}\PY{n}{mu}\PY{p}{,} \PY{n+nb+bp}{self}\PY{o}{.}\PY{n}{r}\PY{p}{,} \PY{n+nb+bp}{self}\PY{o}{.}\PY{n}{v}\PY{p}{)}
\end{Verbatim}
We compute the following orbital elements for the orbit of the ISS:%
\begin{itemize}

\item 

Semi-major axis: $a=6777.773$ km
\item 

Eccentricity: $e=0.00109$
\item 

Inclination: $i=51.724$ deg
\item 

Longitude of ascending node: $\Omega=82.803$ deg
\item 

Argument of periapsis: $\omega=101.293$ deg
\item 

True anomaly: $\nu=48.984$ deg
\end{itemize}

Based on the orbital elements derived quantities like the orbital period can be determined.

In the idealized two-body problem which assumes a uniform gravity potential the only orbital element that changes over time is the true anomaly.
It is the angle that defines the position of the spacecraft on the orbital ellipse.
By solving Kepler's equation we can determine the true anomaly for every point in time and derive new Cartesian state vectors \cite{DAV13}.\begin{Verbatim}[commandchars=\\\{\},fontsize=\footnotesize]
\PY{n}{kepler\PYZus{}orbit} \PY{o}{=} \PY{n}{iss}\PY{o}{.}\PY{n}{kepler\PYZus{}orbit}\PY{p}{(}\PY{p}{)}
\PY{n}{kepler\PYZus{}orbit}\PY{o}{.}\PY{n}{plot3d}\PY{p}{(}\PY{p}{)}
\end{Verbatim}
We now call the \texttt{kepler\_orbit} instance method to solve Kepler's equation at regular intervals until one revolution is completed.
The trajectory that comprises of the resulting state vectors is stored in the returned \texttt{Orbit} object.
By calling \texttt{plot3d} we receive a three-dimensional visualization\DUfootnotemark{id20}{id21}{3} of the full orbital ellipse as shown in figure \DUrole{ref}{3d}.%
\DUfootnotetext{id21}{id20}{3}{
The visualization suffers from the fact that Matplotlib's mplot3d toolkit does not support proper depth buffering. Thus Matplotlib renders the complete orbit in front of the Earth and the parts of the orbit behind the planet are not obscured.}
\begin{figure}[]\noindent\makebox[\columnwidth][c]{\includegraphics[width=\columnwidth]{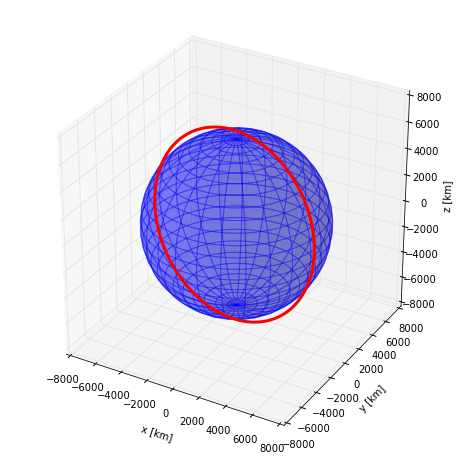}}
\caption{A three-dimensional visualization of the orbit based on Matplotlib. \DUrole{label}{3d}}
\end{figure}

We can achieve a similar result, apart from numerical errors, by numerically integrating Newton's equation:\begin{equation}
\label{newton}
\vec{\ddot{r}} = -\mu \frac{\vec{r}}{|\vec{r}|^3}
\end{equation}Plyades uses the DOP853 integrator from the \texttt{scipy.integrate} suite which is an 8th-order Runge-Kutta integrator with Dormand-Prince coefficients.
By default the propagator uses adaptive step-size control and a simple force model that only considers the uniform gravity potential (see equation \DUrole{ref}{newton}).\begin{Verbatim}[commandchars=\\\{\},fontsize=\footnotesize]
\PY{n}{newton\PYZus{}orbit} \PY{o}{=} \PY{n}{iss}\PY{o}{.}\PY{n}{propagate}\PY{p}{(}
    \PY{n}{iss}\PY{o}{.}\PY{n}{period}\PY{o}{*}\PY{l+m+mf}{0.8}\PY{p}{,}
    \PY{n}{max\PYZus{}step}\PY{o}{=}\PY{l+m+mi}{500}\PY{p}{,}
    \PY{n}{interpolate}\PY{o}{=}\PY{l+m+mi}{100}
\PY{p}{)}
\PY{n}{newton\PYZus{}orbit}\PY{o}{.}\PY{n}{plot\PYZus{}plane}\PY{p}{(}\PY{n}{plane}\PY{o}{=}\PY{l+s+s1}{\PYZsq{}}\PY{l+s+s1}{XZ}\PY{l+s+s1}{\PYZsq{}}\PY{p}{,} \PY{n}{show\PYZus{}steps}\PY{o}{=}\PY{n+nb+bp}{True}\PY{p}{)}
\end{Verbatim}
In this example we propagate for 0.8 revolutions and constrain the step size to 500 seconds to improve accuracy.
We also interpolate additional state vectors between the integrator steps for visualization purposes.\begin{figure}[]\noindent\makebox[\columnwidth][c]{\includegraphics[width=\columnwidth]{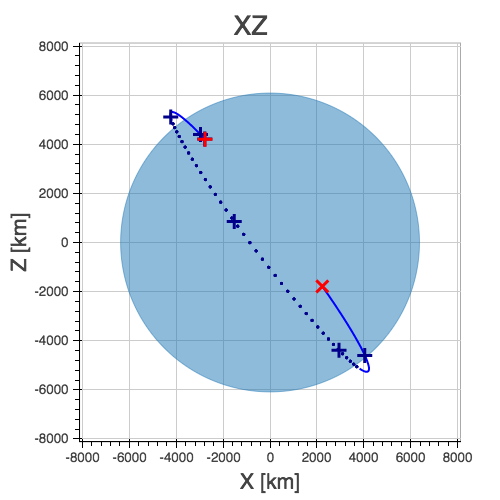}}
\caption{Visualization of a numerically propagated orbit with intermediate solver steps (+, blue), start point (+, red), and end point (x, red). \DUrole{label}{numerical}}
\end{figure}

The trajectory plot in figure \DUrole{ref}{numerical} also includes markers for the intermediate integrator steps.

Since the shape of the Earth is rather an irregular ellipsoid than a sphere Earth's gravity potential is also not uniform.
We can model the oblateness of the Earth by including the second dynamic form factor $J_2$ in the equations of motion as shown in equation \DUrole{ref}{j2}.\begin{equation}
\label{j2}
    \vec{\ddot{r}} = -\mu \frac{\vec{r}}{|\vec{r}|^3} - \frac{3}{2} \frac{\mu J_2 R_e^2}{|\vec{r}|^5} \begin{bmatrix} x \left(1 - 5\frac{z^2}{|\vec{r}|^2}\right) \\ y \left(1 - 5\frac{z^2}{|\vec{r}|^2}\right) \\ z \left(3 - 5\frac{z^2}{|\vec{r}|^2}\right) \end{bmatrix}
\end{equation}When introducing this perturbation we should expect that the properties of the orbit will change over time.
We will now analyze these effects further.

Plyades allows the substitution of force equations with a convenient decorator-based syntax that is illustrated in the next code listing.\begin{Verbatim}[commandchars=\\\{\},fontsize=\footnotesize]
\PY{n+nd}{@iss.gravity}
\PY{k}{def} \PY{n+nf}{newton\PYZus{}j2}\PY{p}{(}\PY{n}{f}\PY{p}{,} \PY{n}{t}\PY{p}{,} \PY{n}{y}\PY{p}{,} \PY{n}{params}\PY{p}{)}\PY{p}{:}
    \PY{n}{r} \PY{o}{=} \PY{n}{np}\PY{o}{.}\PY{n}{sqrt}\PY{p}{(}\PY{n}{np}\PY{o}{.}\PY{n}{square}\PY{p}{(}\PY{n}{y}\PY{p}{[}\PY{p}{:}\PY{l+m+mi}{3}\PY{p}{]}\PY{p}{)}\PY{o}{.}\PY{n}{sum}\PY{p}{(}\PY{p}{)}\PY{p}{)}
    \PY{n}{mu} \PY{o}{=} \PY{n}{params}\PY{p}{[}\PY{l+s+s1}{\PYZsq{}}\PY{l+s+s1}{body}\PY{l+s+s1}{\PYZsq{}}\PY{p}{]}\PY{o}{.}\PY{n}{mu}\PY{o}{.}\PY{n}{value}
    \PY{n}{j2} \PY{o}{=} \PY{n}{params}\PY{p}{[}\PY{l+s+s1}{\PYZsq{}}\PY{l+s+s1}{body}\PY{l+s+s1}{\PYZsq{}}\PY{p}{]}\PY{o}{.}\PY{n}{j2}
    \PY{n}{r\PYZus{}m} \PY{o}{=} \PY{n}{params}\PY{p}{[}\PY{l+s+s1}{\PYZsq{}}\PY{l+s+s1}{body}\PY{l+s+s1}{\PYZsq{}}\PY{p}{]}\PY{o}{.}\PY{n}{mean\PYZus{}radius}\PY{o}{.}\PY{n}{value}
    \PY{n}{rx}\PY{p}{,} \PY{n}{ry}\PY{p}{,} \PY{n}{rz} \PY{o}{=} \PY{n}{y}\PY{p}{[}\PY{p}{:}\PY{l+m+mi}{3}\PY{p}{]}
    \PY{n}{f}\PY{p}{[}\PY{p}{:}\PY{l+m+mi}{3}\PY{p}{]} \PY{o}{+}\PY{o}{=} \PY{n}{y}\PY{p}{[}\PY{l+m+mi}{3}\PY{p}{:}\PY{p}{]}
    \PY{n}{pj} \PY{o}{=} \PY{o}{\PYZhy{}}\PY{l+m+mi}{3}\PY{o}{/}\PY{l+m+mi}{2}\PY{o}{*}\PY{n}{mu}\PY{o}{*}\PY{n}{j2}\PY{o}{*}\PY{n}{r\PYZus{}m}\PY{o}{*}\PY{o}{*}\PY{l+m+mi}{2}\PY{o}{/}\PY{n}{r}\PY{o}{*}\PY{o}{*}\PY{l+m+mi}{5}
    \PY{n}{f}\PY{p}{[}\PY{l+m+mi}{3}\PY{p}{]} \PY{o}{+}\PY{o}{=} \PY{o}{\PYZhy{}}\PY{n}{mu}\PY{o}{*}\PY{n}{rx}\PY{o}{/}\PY{n}{r}\PY{o}{*}\PY{o}{*}\PY{l+m+mi}{3} \PY{o}{+} \PY{n}{pj}\PY{o}{*}\PY{n}{rx}\PY{o}{*}\PY{p}{(}\PY{l+m+mi}{1}\PY{o}{\PYZhy{}}\PY{l+m+mi}{5}\PY{o}{*}\PY{n}{rz}\PY{o}{*}\PY{o}{*}\PY{l+m+mi}{2}\PY{o}{/}\PY{n}{r}\PY{o}{*}\PY{o}{*}\PY{l+m+mi}{2}\PY{p}{)}
    \PY{n}{f}\PY{p}{[}\PY{l+m+mi}{4}\PY{p}{]} \PY{o}{+}\PY{o}{=} \PY{o}{\PYZhy{}}\PY{n}{mu}\PY{o}{*}\PY{n}{ry}\PY{o}{/}\PY{n}{r}\PY{o}{*}\PY{o}{*}\PY{l+m+mi}{3} \PY{o}{+} \PY{n}{pj}\PY{o}{*}\PY{n}{ry}\PY{o}{*}\PY{p}{(}\PY{l+m+mi}{1}\PY{o}{\PYZhy{}}\PY{l+m+mi}{5}\PY{o}{*}\PY{n}{rz}\PY{o}{*}\PY{o}{*}\PY{l+m+mi}{2}\PY{o}{/}\PY{n}{r}\PY{o}{*}\PY{o}{*}\PY{l+m+mi}{2}\PY{p}{)}
    \PY{n}{f}\PY{p}{[}\PY{l+m+mi}{5}\PY{p}{]} \PY{o}{+}\PY{o}{=} \PY{o}{\PYZhy{}}\PY{n}{mu}\PY{o}{*}\PY{n}{rz}\PY{o}{/}\PY{n}{r}\PY{o}{*}\PY{o}{*}\PY{l+m+mi}{3} \PY{o}{+} \PY{n}{pj}\PY{o}{*}\PY{n}{rz}\PY{o}{*}\PY{p}{(}\PY{l+m+mi}{3}\PY{o}{\PYZhy{}}\PY{l+m+mi}{5}\PY{o}{*}\PY{n}{rz}\PY{o}{*}\PY{o}{*}\PY{l+m+mi}{2}\PY{o}{/}\PY{n}{r}\PY{o}{*}\PY{o}{*}\PY{l+m+mi}{2}\PY{p}{)}
\end{Verbatim}
\begin{figure}[]\noindent\makebox[\columnwidth][c]{\includegraphics[width=\columnwidth]{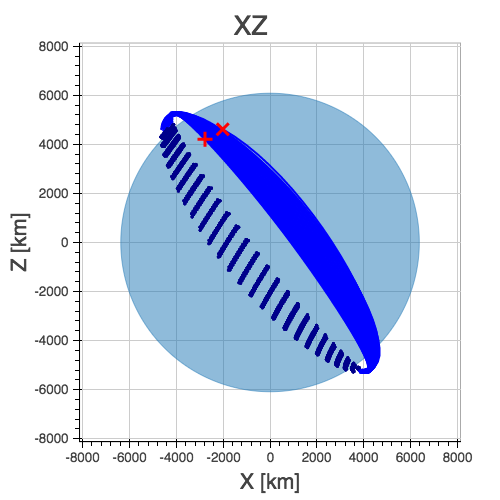}}
\caption{Visualization of the perturbed orbit. \DUrole{label}{perturbed}}
\end{figure}After propagating over 50 revolutions the perturbation of the orbit is clearly visible within the visualization in figure \DUrole{ref}{perturbed}.
A secular (non-periodical) precession of the orbital plane is visible.
Thus a change in the longitude of the ascending node should be present.

We can plot the longitude of the ascending node by issuing the following command:\begin{Verbatim}[commandchars=\\\{\},fontsize=\footnotesize]
\PY{n}{j2\PYZus{}orbit}\PY{o}{.}\PY{n}{plot\PYZus{}element}\PY{p}{(}\PY{l+s+s1}{\PYZsq{}}\PY{l+s+s1}{ascending\PYZus{}node}\PY{l+s+s1}{\PYZsq{}}\PY{p}{)}
\end{Verbatim}
The resulting figure \DUrole{ref}{osculating} shows the expected secular change of the longitude of the ascending node.\begin{figure}[]\noindent\makebox[\columnwidth][c]{\includegraphics[scale=0.40]{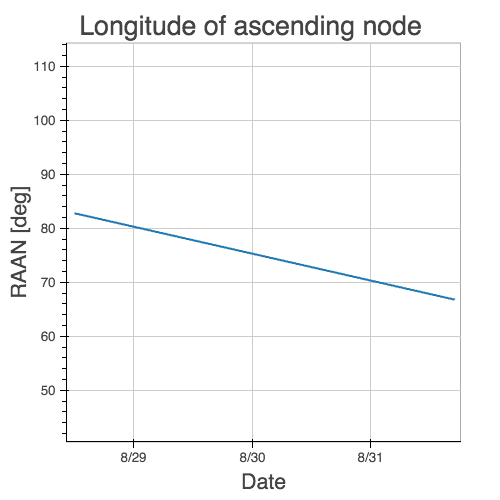}}
\caption{Secular perturbation on the longitude of the ascending node. \DUrole{label}{osculating}}
\end{figure}

\section{Future Development%
  \label{future-development}%
}

As of this writing Plyades has been superseded by the Python Astrodynamics project \cite{PyA15}.
The project aims to merge the three MIT-licensed, Python-based astrodynamics libraries Plyades, Poliastro \cite{JCR15} and Orbital \cite{FML15} and provide a comprehensive Python-based astrodynamics toolkit for productive use.

\section{Conclusion%
  \label{conclusion}%
}

In this paper we have discussed the current tools and programming environments for space mission design.
These suffer from high complexity, low levels of abstraction, low flexibility, and out-dated programming practices.
We have then shown why the maturity and breadth of the scientific Python ecosystem as well as the usability of the Python programming language make Python a viable alternative for next generation astrodynamics tools.
With the design and implementation of the proof-of-concept library Plyades we demonstrated that it is possible to create powerful yet simple to use astrodynamics tools in pure Python by using scientific Python libraries and following modern best practices.
The Plyades work has lead to the foundation of the Python Astrodynamics project, an inter-european collaboration, whose goal is the development of a production-grade Python-based astrodynamics library.


\end{document}